# Polarization-sensitive mechanically-tunable microwave filter using metallic photonic crystals

Sanaz Zarei

*Abstract*— A tunable, polarization sensitive microwave filter based on a reconfigurable dual-layered two-dimensional metallic photonic crystal is presented. The filter consists of a metallic plate with periodic holes and metal pillars which are anchored in a substrate material. The diameter of the pillars is smaller than that of the holes and the lateral position of the pillars with respect to the holes is tunable. We study the transmission of the device for linearly polarized electromagnetic radiation, which is polarized parallel and perpendicular to the displacement of the movable membrane, respectively. The tuning range spans from 20.18 GHz to 26.58 GHz for perpendicular polarization and from 20.18 GHz to 18.42 GHz for parallel polarization. The 3 dB operation bandwidth varies between 3.82 GHz and 6.44 GHz for perpendicular polarization and between 3.42 GHz and 3.82 GHz for parallel polarization. Experimental data are in good agreement with finite element method (FEM) simulations.

## I. INTRODUCTION

TUNABLE microwave filters are essential components in systems for a wide range of applications to remove undesired signals. This includes radar systems, broadband wireless networks and satellite communications [1]. Over the years, several concepts for tunable microwave filters based on photonic crystals [2-10], metamaterials [11-15], and frequency selective surfaces (FSS) [16-24] have been proposed. While photonic crystals are periodic structures of dielectrics with different refractive indices, metamaterials are composed of periodically subwavelength metal/dielectric structures. Frequency selective surfaces are a special case of the metasurfaces, i.e., 2D metamaterials. These classes of artificial materials have been explored extensively as they offer the possibility of creating novel electromagnetic properties that are not available in naturally existing materials in microwave, terahertz and optical frequency bands. The main benefit of these artificial structures is their ability to control and manipulate electromagnetic waves.

Among several reported tunable filters that are working in the microwave region [2-26], tunable microwave photonic filter using photonic crystal nanocavities with tunable central frequency from 13 to 20GHz [5], tunable notch microwave photonic filter based on photonic crystal nanocavity with central frequency tuning of 12.9-32.3GHz [7], 4D-printed origami-inspired tunable multi-layer FSS with tunable resonant frequency from 22.4GHz to 26.1GHz [17], dual-polarization absorptive/transmissive FSS with tunable passband from 3.56 to 4.35GHz [21], and tunable FSS based on a sliding 3D-printed inserted dielectric with a continuous tunable range of 3.61 to 5.89GHz [23], can be named.

In this letter, a polarization sensitive tunable filter based on extraordinary transmission through subwavelength annular apertures in a dual-layered metallic structure is fabricated and characterized in the microwave frequency range. In a previous work, theoretical demonstration of a similar tunable filter in the terahertz frequency range has been accomplished [27]. In this work, both simulations and experimental results show that the resonance frequency of the filter can be tuned easily by changing the lateral spacing between the holes on the upper metal plate and the inner pillars that are attached to the substrate (see Fig. 1).

## II. FILTER DESIGN AND FABRICATION

Figure 1 shows a schematic of the reconfigurable dual-layered array of subwavelength metallic pillar-hole structure. A brass-colored plate can be seen at the top, which represents the metal plate with an array of holes. Pins project into these holes and are anchored on a dielectric plate (blue). The two plates can be shifted against each other so that the pins can be centered in the holes on the one hand, but can also be shifted within the holes. Fig. 1(a) depicts its initial state, i.e., when pillars are in the center of the holes. Fig. 1(b) illustrates how the device is tuned by a lateral displacement of holes with respect to the pillars, along either the x- or the y-direction. The displacement is defined as the displacement of the pillars from the center of the holes.

The important parameters in the filter design are the cell dimension ($\Lambda$), inner radius ($r_i$), outer radius ($r_o$), and the thickness of the metal plate ($H_1$). Pillars are placed in a way that their top surface and the metal plate top surface are in the same plane. The structure dimensions were set to $\Lambda = 10.81$ mm, $r_i = 2.00$ mm, $r_o = 3.25$ mm, and $H_1 = 5.15$ mm, since a filter with such a design can be manufactured easily. The transmission through the structure was numerically analyzed for two perpendicular polarizations of linearly polarized incident radiation, which was propagating in the z-direction in Fig. 1 (along the pillars). One polarization was considered perpendicular to the moving axis of the metal plate and the other was considered parallel to the moving axis and are referred to as perpendicular and parallel polarization in this letter, respectively.

Sanaz Zarei was with the School of Electrical and Computer Engineering, University of Tehran, Tehran, Iran (e-mail: szarei@ut.ac.ir).

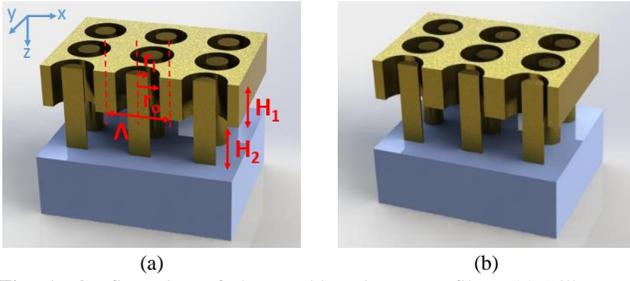

**Fig. 1.** Configuration of the tunable microwave filter. (a) Pillars are centered inside the holes, (b) The metal plate perforated by holes is displaced laterally with respect to pillars.

The electromagnetic modeling of the structure was performed using the software COMSOL. To achieve practical simulation times, it was necessary to model the device as an infinite grid by surrounding the unit cell of the device with periodic boundary conditions. The pillars and the metal plate were assumed to be perfect electric conductors, which is a good approximation at gigahertz frequencies, due to the negligible losses of metals. In the initial simulations, a dielectric substrate holding the pillars was not considered. In addition, the pillars were assumed to be of the same length as the metal plate thickness.

The filter with the abovementioned design was fabricated by the mechanical workshop. First, the holes were drilled periodically into an aluminum plate. To obtain the precise thickness, a thicker metal plate was selected and was polished until the desired thickness was achieved. The copper pillars were made longer than the thickness of the metal plate and were anchored in a substrate plate made from polypropylene. The pillars were anchored into the substrate with the same periodicity as the holes in the metal plate. The total length of the pillars was 19.00 mm and the substrate thickness was 10.00 mm. The pillars depth within the substrate was 8.50 mm. The filter size is 10×10 cm² and consists of an array of 9×9 ring apertures. A photo of the fabricated filter is shown in Fig. 2.

### III. DESCRIPTION OF THE MEASUREMENT METHOD

The metal plate was fixed on an adapter and mounted on a manual linear translation stage which allows for a controlled movement in the horizontal direction. The substrate plate with pillars was also fixed on an adapter. The two adapters could be moved in the vertical direction and rotated along the vertical axis. They were aligned parallel to each other with a help of a spacer, which also insured that the surfaces of the metal plate and the pillars were in the same plane. Additionally, the spacer also enabled easier positioning of the pillars in the middle of the holes before fixing the adapter position. Once aligned, the filter was positioned between the microwave emitter and detector and a metal shielding was built around it, which prevented stray radiation from reaching the detector. After the transmission measurements had been performed, the filter was removed and the reference measurement was taken. Then the emitter and the detector were turned by 90 degrees to switch from parallel to perpendicular polarization of incident waves and the procedure was repeated.

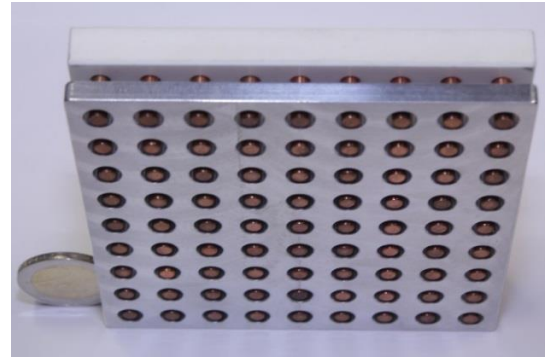

**Fig. 2.** Photo of the fabricated microwave filter. Note the two-euro coin for size comparison.

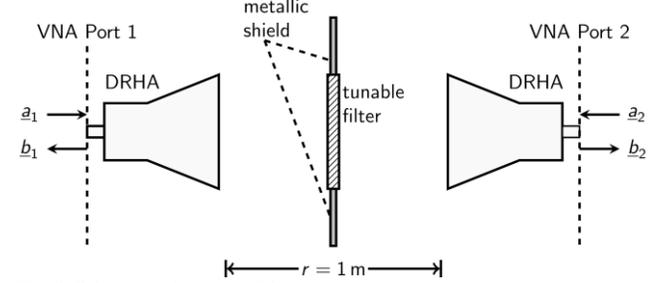

**Fig. 2.** Schematic drawing of the measurement setup.

The filter transmission coefficient was measured in a free-space setup (see Fig. 3). For this purpose, two double ridged horn antennas (DRHA, Type: RFSpin DRH50) were aligned properly at a distance of 1 m from each other and the filter to be characterized was placed in the middle between both antennas. To hold the filter, a metallic shield was used, which guarantees that only the transmission through the filter was measured.

Unfortunately, the transmission coefficient of the filter cannot be measured directly since the reference planes of the measured scattering parameters are at the feeding points of the antennas. The filter's two-port is thus embedded by two surrounding two-ports. Assuming a plane wave at the location of the filter and neglecting reflections from the antennas, the absolute value of the transmission coefficient can be determined with the aid of a reference measurement:

$$\left| S_{21}^{filter} \right| \approx \frac{\left| S_{21}^{filled} \right|}{\left| S_{21}^{empty} \right|} \quad (1)$$

with $\left| S_{21}^{empty} \right|$ being the measured transmission coefficient of the full system without filter and $\left| S_{21}^{filled} \right|$ being the measured transmission coefficient of the full system with filter. During both measurements, radio frequency absorbers have been placed next to the antennas to suppress reflections. Furthermore, the measured transmission data for each displacement of the metal plate was smoothed applying a cubic Savitzky-Golay filter with data frames (window sizes) of 0.41 GHz, which also reduces the influence of unwanted reflections.

### IV. RESULTS

The initially-simulated transmission of the designed microwave filter is shown in figures 4(a) and 4(b). The zero displacement corresponds to the pillars positioned in the middle of the holes. The holes can be displaced into both directions on a moving axis. Only three displacements in a single direction and the zero displacement are shown for clarity. The displacement in the opposite direction resulted in the same filter response as expected due to the symmetry of the filter design. For zero displacement we observe a distinct transmission resonance around 19 GHz with a FWHM of 1.46 GHz resulting in a Q-factor of 13.45. The resonance shifts as the pillars are displaced with respect to the holes. For perpendicular polarization (Fig. 4(a)) the resonance shifts to higher frequencies. In contrast, the transmission peak shifts to lower frequencies (Fig. 4(b)) for parallel polarized waves.

These numerical results, in principle, are confirmed by the experiments which are shown in figures 4(c) and 4(d). The experimental data look somewhat noisy, which is typical when characterizing microwave components with free-space optics. We attribute this noise, which can be reduced by a Savitzky-Golay filter, to interference effects. The measured peak transmission is lower than the simulated one for both polarizations. We attribute this to the additional length of the pillars as well as the reflection and absorption losses in the substrate. Although the measurement results do not match the numerical simulations one-to-one, they reproduce the main trend clearly. For a better comparison, the simulated transmission characteristics of the fabricated filter are also plotted in figures 4(e) and 4(f). Further optimization of the substrate thickness, total length of the pillars, and $H_2$ can lead to a better matching between Initial-simulation and experimental results.

The peak-transmission frequency and the 3 dB bandwidth obtained from simulations and measurements are shown in Fig. 5. The simulation results show that the tuning range for perpendicular polarization ranges from 19.24 GHz to 26.08 GHz. The 3 dB bandwidth reduces from 1.44 GHz to 1.34 GHz when the displacement is increased (Fig. 5(a)). For parallel polarization, the tuning range extends from 19.24 GHz to 15.82 GHz while the 3 dB bandwidth decreases from 1.44 GHz to 0.82 GHz by increasing the displacement (Fig. 5(b)). This gives an approximate tuning sensitivity of 6.84 GHz per 1.25mm displacement (i.e., 5.47 GHz∕mm) for perpendicular polarization and 3.42 GHz per 1.25mm displacement (i.e., 2.74 GHz / mm) for parallel polarization. The corresponding Q-factor is 13.6, rising to 19.3 for both polarizations by increasing the displacement. The experimental results show a similar behavior except for the 3 dB bandwidth for perpendicularly polarized waves. The experimentally obtained tuning range for perpendicular polarization ranges from 20.18 GHz to 26.58 GHz. The 3 dB bandwidth increases from 3.82 GHz to 6.44 GHz (Fig. 5(c)). The tuning range fits well with the simulated results, but the 3 dB bandwidth shows a different trend compared to the simulation. For parallel polarization the tuning range ranges from 20.18 GHz to 18.45 GHz. The 3 dB bandwidth decreases from 3.82 GHz to 3.42 GHz when increasing the displacement (Fig. 5(d)). This indicates a smaller tuning range and a bigger overall 3dB bandwidth compared to the simulation results.

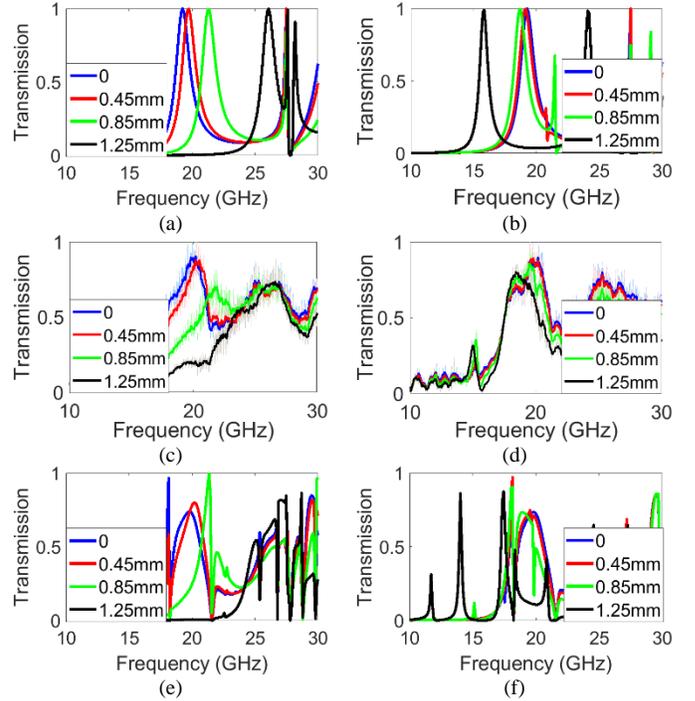

**Fig. 4.** (a) and (b) The initially-simulated transmission of the designed tunable microwave filter, (c) and (d) the measured transmission of the fabricated microwave filter, and (e) and (f) the simulated transmission characteristics of the fabricated microwave filter. (a), (c) and (e) For incident waves linearly polarized perpendicular to the moving axis and (b), (d) and (f) for incident waves linearly polarized along the moving axis. The thick curves shown in (c) and (d) are smoothed recorded data using Savitzky-Golay filter and the original data are shown in the background as shaded lines. The blue curves show the device transmission in initial state and the red, green and black curves show the transmission for the lateral displacement of 0.45, 0.85 and 1.25mm, respectively.

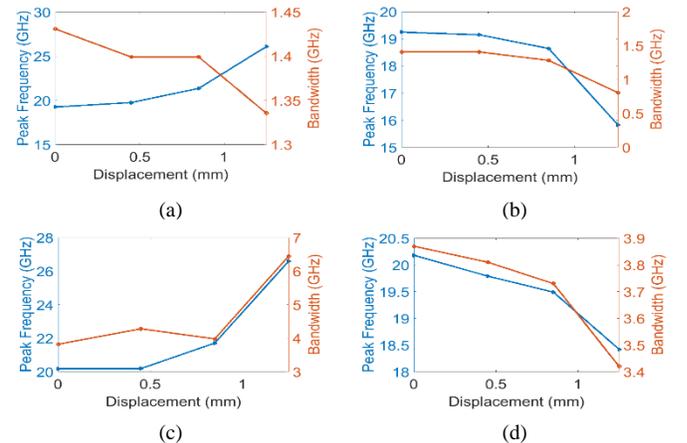

**Fig. 5.** Initially-simulated transmission-peak frequency and 3 dB bandwidth as a function of displacement for (a) perpendicular and (b) parallel polarization. Experimental transmission-peak frequency and 3 dB bandwidth in terms of displacement for (c) perpendicular and (d) parallel polarization.

The approximate tuning sensitivities of the filter are 6.4 GHz per 1.25mm displacement (i.e., 5.12 GHz∕mm) and 1.76 GHz per 1.25mm displacement (i.e., 1.41 GHz / mm) for perpendicular and parallel polarizations, respectively. For perpendicular polarization the Q-factor is approximately 5.3. It reduces to 4.1 by increasing the displacement. For parallel polarization it is nearly 5.3 for the entire tuning range.

## V. Discussion

### A. Polarization Response

As is stated before, the presented filter is polarization sensitive. When the polarization is perpendicular to the moving axes of the membrane, the resonance frequency of the device blue-shifts with increasing the membrane's displacement, while for the polarization parallel to the moving axes, the resonance frequency red-shifts by the displacement.

### B. Frequency Tuning Range

The measured frequency tuning range of the device for perpendicular polarization is 6.4GHz, while it is 1.76GHz for parallel polarization. Therefore, the filter has a wider frequency tuning range for perpendicular polarization and thus a better performance. Also, as the geometrical parameters of the metallic array dramatically affect the response of the filter, these can be optimized via a design space exploration to determine the maximum tuning range at any desired frequency.

### C. Tuning Sensitivity

The tuning sensitivity of the filter is not uniform through the whole tuning range for both polarizations.

### D. Tuning Mechanism

The physical mechanism of the proposed tunable bandpass filter is explained from the perspective of propagation constant of the mode involved in power transmission in [27].

### E. Device Scalability

Since the transmission properties strongly depend on the geometrical parameters of the metallic hole array, it is possible to tune the transmission pass-band at other frequency ranges by appropriately choosing the dimensions. For instance, the transmission pass-band frequency decreases by increasing the periodicity, membrane thickness, inner and outer radii of the rings, while the transmission bandwidth decreases by increasing the periodicity, thickness of metallic membrane, inner radius of the rings and increases by increasing the outer radius of the rings. Thereby, it is possible to optimize the filter performance at desired frequencies.

### F. Structure Symmetry

Due to the symmetry of metallic array, it is sufficient to present the results for only a one lateral moving axis (let's say x-axis) and just for a number of representative lateral displacements on a half of the rings' diameter (instead of the whole diameter).

### G. Switching Performance

The device presented in this study has the capability of working as a digital device, with either the ON-state (when the pins are centered inside the holes) and OFF-state (when the metallic membrane is in contact with the pins).

### H. Reflection-mode Performance

The presented tunable filter works in the transmission mode. However, a reflection-type filter with a tunable stop-band can also be available based on our design.

## VI. Conclusion

In conclusion, a polarization sensitive tunable microwave filter based on extraordinary transmission through subwavelength ring-shaped apertures in a dual-layered metallic structure was fabricated and characterized in this letter. Both FEM simulations and experimental results show that the resonance frequency of the filter can be reconfigured easily by varying the lateral spacing between the holes on the upper metal plate and the inner pillars that are attached to the substrate. The measurements indicate a tuning range from 20.18 GHz to 26.58 GHz for the waves linearly polarized perpendicular to the moving axes of the metal plate with the 3 dB bandwidth varying between 3.82 GHz and 6.44 GHz. If the electromagnetic waves are linearly polarized parallel to the moving axes, we determine a tuning range from 20.18 GHz to 18.42 GHz with a 3 dB bandwidth of 3.82 GHz to 3.42 GHz.